\documentstyle[12pt]{article}
\topmargin - 1.5cm
\leftmargin - 1cm
\begin{document}

\begin{center}

{\large \bf On the qualitative theory of non-Gamov decay states\\[1cm]}
{\sc V. M. Chabanov and B. N. Zakhariev 
\\[3mm]} 
Bogoliubov Laboratory of Theoretical Physics, Joint Institute for 
Nuclear Research, \\ 141980 Dubna, Russia \\ 
e-mail:  zakharev@thsun1.jinr.ru;  chabanov@thsun1.jinr.ru

\end{center}

\begin{center}
Abstract
\end{center}

Complex potential transformations which add imaginary parts to
chosen energy levels are given and qualitatively explained.  
Unexpected shape similarity of potential perturbations 
for real and imaginary E-shifts of bound states are exhibited. 
The imaginary E-shifts in the continuous spectrum lead to a 
surprising quasi-periodic field raking up initial propagating 
waves into localized states. Complex periodic potentials without 
lacunas (!) are constructed.  The fission of quasi-bound states when 
neighbour complex eigenvalues approach one another is 
demonstrated.  E-shift algorithms represent  wide classes of  exactly 
solvable quantum models for non-self-adjoint operators.  

\vspace{0.4cm}

\section{Introduction}

The significant success  has been  achieved last years due to 
inverse problem (IP) and SUSYQ approach [1-7]. 
The merit of IP is that
the observables are input data in this formalism and their variations 
correspond to a complete set of IP exactly solvable models (ESM).  This 
inspires an idea  to 
control spectral parameters, which may allow one to unveil new
relations between the spectral, scattering, decay data 
and forces acting in quantum objects, see examples in 
[8-12].  Such a concept has been realized last years 
when we have developed a qualitative theory of quantum design 
[5,6,13-17].  Through the computer visualization, we have 
established physically clear algorithms of elementary transformations 
associated with given variations of distinct spectral parameters.  We have 
succeeded in revealing  universal potential blocks from which any quantum 
systems may be engineered, at least mentally.  All this is by no means 
reduced only to content of the distinct images but 
embraces continuum of all the elementary variations 
 of potential shapes. So one acquires a "mathematical vision" in 
microworld which is hidden from our eyes.                             

In this paper we consider the generalization of quantum design to imaginary 
shifts of energy values leading to non-self-adjoint Schr\"odinger
operator (complex-valued potentials). In the mathematical physics,
non-Hermitian operators represent an area of permanent interest. 
The spectral analysis of non-Hermitian differential operators 
was considered by Lyance V.E. in Appendix 1 to the 2nd ed. of 
the  book by Naimark \cite{NL}.  
As a widely known example of applying complex potentials in 
physics, the optical model can be given. This is an effective tool
for accounting the processes with particle number nonconservation  in
a specific channel occurring in many nuclear (e. g. heavy-ion)  reactions.
There were some  papers devoted to  construction of complex potentials with 
pure real spectrum of eigenvalues, see \cite{Zn} and references 
therein.  The SUSYQ approach was applied to generate complex potential 
partners with the same eigenvalues  spectra distinguishing in only
one  eigenenergy (or finite, but given, number of spectrum points), see
\cite{B,A}. Here we present  a technique of shifting energy
levels into complex energy
plane. This makes more complete the set of ESM for non-self-adjoint 
Schr\"odinger operators.

\section{Formalism}

To add imaginary value to the energy of chosen state
we shall deal with SUSYQ approach, see, papers 
\cite{CKS,Suk,B,A}. However, we perform SUSYQ transformation
twice at {\it different} (and complex) energies.  In the first step, the 
SUSYQ procedure results in deletion of state at initial energy level. The 
second transformation is made at the energy level whose 
real part coincides with initial energy value whereas the imaginary 
value is added.  As a result a chosen state energy is shifted in a complex 
"direction", at that all the remaining energy levels remain unaltered.  
This is extension of the results on only creation (or deletion) 
of complex energy eigenvalue (SUSYQ transformation of 
non-Hermitian operators). 

 In the SUSYQ approach the initial 
Hamiltonian $H_{0}$, the second-order differential operator, is factorized
into first order operators 
\begin{eqnarray}
A^{\pm} \enskip = \enskip \pm \partial + W(x), 
\label{w}
\end{eqnarray}
where $\partial$ is a symbol of the derivative :
\begin{eqnarray} 
H_{0} \enskip = \enskip A^{+} A^{-} + {\cal E}. 
\label{H0}
\end{eqnarray} 
Symbol ${\cal E}$ denotes a constant factorization energy and
superpotential $W(x)$ is determined by the equation
\begin{eqnarray} 
A^{-} \psi_{0}(x,{\cal E}) \enskip = \enskip 
\{-\partial + W(x) \} \psi_{0}(x,{\cal E}) \enskip = \enskip 0.
\label{weq}
\end{eqnarray} 
Here $\psi_{0}(x,{\cal E})$ is the solution of Schr\"odinger equation
with the initial potential $V_{0}(x)$ at the factorization energy ${\cal 
E}$.  We have from Eqs. (\ref{w}) and (\ref{weq}) 
\begin{eqnarray} 
W(x) \enskip = \enskip \psi_{0}'(x,{\cal E}) \psi_{0}(x,{\cal E})^{-1}.  
\label{w1}
\end{eqnarray} 

The  SUSYQ transformation consists in permutation of $A^{\pm}$ :  
\begin{eqnarray} H_{0} \enskip = \enskip A^{+} 
A^{-} + {\cal E} \rightarrow H_{1} \enskip = \enskip A^{-} A^{+} 
+ {\cal E}; \\
V_{1}(x) \enskip = \enskip V_{0}(x) - 2 W'(x).
\label{trans} 
\end{eqnarray} There is a simple relation  between the 
solutions $\psi_{0}(x,E)$ and $\psi_{1}(x,E)$ of the Schr\"odinger equation 
at arbitrary energy $E$ with the initial and new Hamiltonians 
$H_{0},\,H_{1}$ 
\begin{eqnarray} 
\psi_{1}(x,E)=A^{-} \psi_{0}(x,E)= \nonumber \\
=(-\partial + W(x)) 
\psi_{0}(x,E)  =  \psi_{0}(x,{\cal E})^{-1} \theta_{E} (x),  
\nonumber \\  \theta_{E} (x) \enskip \equiv \enskip \psi_{0}'(x,{\cal 
E}) \psi_{0}(x,E) \nonumber \\ - \psi_{0}(x,{\cal E}) \psi_{0}'(x,E),  
\label{psim} 
\end{eqnarray} where we used  Eq. (\ref{w1}).  New solution at 
${\cal E}$ is \begin{eqnarray} \psi_{1} (x,{\cal E}) \enskip = \enskip 
\psi_{0}(x,{\cal E})^{-1}.  
\label{psin} 
\end{eqnarray} The final results 
of this transformation depend on the choice of ${\cal E}$ and 
$\psi_{0}(x,{\cal E})$.   
 
Let us perform SUSYQ transformation once more, only as an initial system
we shall take already the transformed one. New 
factorization energy is ${\bar {\cal E}} \enskip = \enskip {\cal E} + i 
\Gamma $.  As an analog of $\psi_{0}(x,{\cal E})$ we must take linearly 
independent counterpart of $\psi_{1}(x,{\cal E})$ but at the energy shifted 
to complex plane
\begin{eqnarray} 
{\tilde \psi}_{1}(x,{\bar {\cal E}}) \enskip =   
\enskip A^{-} {\tilde \psi}_{0}(x,{\bar {\cal E}}).
\label{psinn}
\end{eqnarray} 
Here ${\tilde \psi}_{0}(x,{\bar {\cal E}})$ stands for the solution which
is obtained from linearly independent counterpart of 
$\psi_{0}(x,{\cal E})$ by adding imaginary value to ${\cal E}$.
The second step gives the transformed solution at ${\bar {\cal E}}$ as 
follows
\begin{eqnarray} 
\psi_{2} (x,{\bar {\cal E}}) \enskip = \enskip 
{\tilde \psi}_{1}(x,{\bar {\cal E}})^{-1} \enskip = \enskip
\psi_{0}(x,{\cal E}) \theta_{{\bar {\cal E}}} (x)^{-1},   
\end{eqnarray}   
where the last equality follows from Eq. (\ref{psim}) and  
$\theta_{{\bar {\cal E}}} (x)$ is given by
\begin{eqnarray}
\theta_{{\bar {\cal E}}}(x) \enskip = \enskip \psi_{0}'(x,{\cal E}) 
{\tilde \psi}_{0}(x,{\bar {\cal E}}) - \psi_{0}(x,{\cal E}) 
{\tilde \psi}_{0}'(x,{\bar {\cal E}}).  
\end{eqnarray}
For our scheme to work the modulus $|\theta_{{\bar {\cal E}}}(x)|$ must
be non-zero throughout the range of solution definition, otherwise 
singularities will appear.  Relative to this question, the following 
reasoning can be adduced.  Since $\theta_{{\bar {\cal E}}}(x) = const$ if 
$\Gamma  =  0$, it is clear that, by continuity, $|\theta_{{\bar {\cal 
E}}}(x)|>0$ for $\Gamma$ lying in some neighbourhood of zero.  On the other 
hand, vanishing of $|\theta_{{\bar {\cal E}}}(x)|$  at any fixed $x$ means 
that logarithmic derivatives of $\psi_{0}(x,{\cal E})$ and ${\tilde 
\psi}_{0}(x,{\bar {\cal E}})$ coincide at this point. But this is hardly
feasible in practice because the only free parameter we have is $\Gamma$, 
whereas we must obey two independent requirements. First, 
we need put imaginary part of ${\tilde \psi}_{0}'(x,{\bar {\cal E}})/ 
{\tilde \psi}_{0}(x,{\bar {\cal E}})$ to be zero and, second,  equate 
the real part to $\psi_{0}'(x,{\cal E})/\psi_{0}(x,{\cal E})$.  In 
principle, however, this does not exclude that $|\theta_{{\bar {\cal 
E}}}(x)|$ might accidentally vanish at some $x$ and ${\bar {\cal E}}$.  So, 
a final solution of the problem depends on computer-based calculations. 
                                                                  
The expression for 
superpotential obtained in the second-step SUSYQ transformation is  given 
by
\begin{eqnarray} 
{\tilde W}(x) = {\tilde \psi}_{1}'(x,{\bar {\cal E}}) 
{\tilde \psi}_{1}(x,{\bar {\cal E}})^{-1} 
\nonumber \\  \hspace*{-3cm}=  \{ \psi_{0}(x,{\cal E})^{-1} \theta_{{\bar 
{\cal E}}} (x) \}'  \psi_{0}(x,{\cal E}) \{ \theta_{{\bar {\cal E}}} (x) 
\}^{-1}  \nonumber \\ =  - \psi_{0}'(x,{\cal E}) 
\psi_{0}(x,{\cal E})^{-1} + \theta_{{\bar {\cal E}}}'(x) 
\theta_{{\bar {\cal E}}}(x)^{-1} =  \nonumber \\ - 
\psi_{0}'(x,{\cal E}) \psi_{0}(x,{\cal E})^{-1}   \nonumber \\
+ ({\bar {\cal E}} - {\cal E}) \psi_{0}(x,{\cal E}) 
{\tilde \psi}_{0}(x,{\bar {\cal E}}) 
\theta_{{\bar {\cal E}}}(x)^{-1}, 
\label{wtil} 
\end{eqnarray} 
where the 
prime denotes the differentiation over the coordinate $x$. Furthermore, we 
used here a known identity $\theta_{{\bar {\cal E}}}'(x) \enskip = \enskip 
({\bar {\cal E}} - {\cal E}) \psi_{0}(x,{\cal E}) {\tilde \psi}_{0}(x,{\bar 
{\cal E}})$. From Eqs.  (\ref{wtil}), (\ref{w1}) and (\ref{trans}) we get 
the resulting expression for second-step potential as follows 
\begin{eqnarray} 
V_{2}(x)  =  V_{1}(x) - 2 {\tilde W}'(x) =  V_{0}(x) 
- 2 ({\bar {\cal E}} - {\cal E}) \nonumber \\ \times \{ 
\psi_{0}(x,{\cal E}) {\tilde \psi}_{0}(x,{\bar {\cal E}}) \theta_{{\bar 
{\cal E}}}(x)^{-1} \}'.  
\label{v2} 
\end{eqnarray} 
By using the identity 
$$\theta_{E}(x) \enskip = \enskip (E - {\cal E}) 
\int^{x} \psi_{0}(y,{\cal E}) \psi_{0}(y, E) dy$$ 
 we can give the expression for unnormalized solution at the 
arbitrary energy related to the potential $V_{2}(x)$ :  
\begin{eqnarray} 
\psi_{2}(x,E)  = ({\cal E} - E)^{-1} 
(-\partial + {\tilde W}(x)) \nonumber \\ \times (-\partial + 
W(x)) \psi_{0}(x,E) \nonumber \\  = 
 ({\cal E} - E)^{-1} (-\partial + {\tilde W}(x)) 
\psi_{0}(x,{\cal E})^{-1} \theta_{E} (x)   \nonumber \\ 
= ({\cal E} - E)^{-1} \psi_{0}'(x,{\cal E}) \psi_{0}(x,{\cal E})^{-2} 
\theta_{E} (x) \nonumber \\ + \psi_{0}(x,E) - ({\cal E} - 
E)^{-1} \psi_{0}'(x,{\cal E}) \psi_{0}(x,{\cal E})^{-2}  
\nonumber \\ \times \theta_{E} (x)+ ({\cal E} - E)^{-1} 
{\tilde \psi}_{0}(x,{\bar {\cal 
E}}) \theta_{E}(x) \theta_{{\bar {\cal E}}}(x)^{-1}  \nonumber\\  = 
\psi_{0}(x,E)  + ({\cal E} - E)^{-1}  {\tilde 
\psi}_{0}(x,{\bar {\cal E}})  \nonumber \\  \times \theta_{E}(x) 
\theta_{{\bar {\cal E}}}(x)^{-1} = \psi_{0}(x,E)    - {\tilde 
\psi}_{0}(x,{\bar {\cal E}})  \theta_{{\bar {\cal 
E}}}(x)^{-1} \nonumber \\ 
\times  \int^{x} \psi_{0}(y,{\cal E}) \psi_{0}(y, E) dy, 
\label{psi2} 
\end{eqnarray} 
The results below are based on these formulae  
for both point and continuous spectrum. 

\section{Bound states}

The appearance of imaginary parts of V and E violates the hermiticity 
(norm or flux  are not conserved) which can be interpreted as an effective 
coupling with "hidden channels" as in the optical model.  The wide manifold
of ESM with complete set of spectral parameters allows {\it to control 
absorption} in the constructed quantum systems.

The complex energy shifts require consideration of
 a system of two coupled Schr\"odinger 
equations for $Re \Psi (x)$ and $Im \Psi (x)$ instead of ordinary one. 
A one-dimensional example of the system is given below.
\begin{eqnarray}
-Re \psi'' (x) = (Re E-Re V(x)) Re \psi (x)- \nonumber \\
-(Im E-Im V(x)) Im \psi (x); \nonumber \\ 
-Im \psi'' (x) = (Re E-Re V(x)) Im \psi (x)+ \nonumber\\
+(Im E-Im V(x)) Re \psi (x).  
\label{ReIm} 
\end{eqnarray} 

It was a surprise to reveal that the algorithms of qualitative prediction 
of potential form for eigenvalue imaginary shift are sometimes
the same  as for real energy shifts, although  equations become much 
more complicated.  It 
is evident that an addition of a constant imaginary potential produces a 
parallel imaginary shift of all spectral points. But for imaginary shift of 
a {\it single} spectral point {\it leaving energies of all other 
states unaltered (elementary transformation)} we need to use special shape 
of potential perturbation. One should take into account the different 
sensitivity to potential variations in space of the complete set of 
orthogonal states. In fact, let us consider imaginary shift of ground 
state in different potentials.  The  potential perturbations 
shapes for the energy shifts $\Delta E = - i$ and $\Delta E = - 1$ for the 
ground state of soliton-like potential are demonstrated in 
figure 1. It is interesting that for the  oscillator potential
corresponding picture turns out to be very similar.   The 
ground states are most sensitive to potential perturbations in the middle 
where the probability distribution has maximum.  To compensate the 
influence of the wells  on the remaining spectrum, additional potential 
bumps on both sides are needed \cite{Z}.  Particularly, the perturbed 
soliton-like potential remains reflectionless for waves in continuum 
spectrum. But the fluxes of incoming and outgoing waves become different.  
This gives possibility to control degree of the wave absorption.
 
 Only the  states shifted into complex $E$-plane become quasi-bound 
because for them the time dependent factor $\exp(-iEt)$ with negative 
imaginary part in $E$ is exponentially decreasing in time. The ordinary 
Gamov decaying states have exponential growth at resonance points of 
scattering matrix in complex energy plane. Unlike this, the wave functions 
of quasi-bound states obtained by imaginary shifts of eigenvalues 
 are usually  {\it quadratically integrable} \cite{B}.  The 
absolute values of all other states remain time independent.

    It is interesting that changing the sign of imaginary 
energy shift $\Delta E= i$ results in sign inversion of $Im \Delta V(x)$.  
But in the case of real shifts $\Delta E = \pm 1$ upward or downward
there is no such exact coincidence of $|\Delta V(x)|$. It is  because 
the real shift of the ground state makes spectrum more dense or 
rarifies it, respectively.   Due to the 
coupling of equations (\ref{ReIm}) there appears  comparatively 
small  $Re \Delta V(x)$ beside the main perturbation $Im \Delta V(x)$.
  
  The form of $\Delta Re V(x)$ perturbation  of the first and the 
second (third and fourth etc.) energy levels for imaginary shifts upwards 
or downwards appeared to be of the same qualitative gross-structure (see 
Fig.2) although the shapes of states differ very much from one another.  
For a long time we could not  explain this fact.  
Surprisingly, it turns out that the simple rule we 
have found for shifts of configurational space-localization of bound states 
without moving energy levels is also valid here. The potential  
barrier-well (well-barrier) block pushes to the right (to the left) the 
corresponding wave function bump if their spatial intervals coincide.  
Each of two first states can be considered as having two 
bumps.  In fact, we can conceive the left and the right parts of the ground 
state as two "bumps" (halves).  The shapes of $\Delta Re V(x)$  in figure 2 
consist of two blocks. The left one (barrier-well) deflects the left side 
of the ground state to the right and the right block (well-barrier) pushes 
the right side of the state to the left. 

We can consider the imaginary shift of the lowest 
level as making the energy spectrum something less dense, compare with 
\cite{ZM}.  So it is naturally to suppose that it will be followed by 
spatial compressing ground state to the center in contrast to widening when 
making spectrum more dense \cite{CZBDS}. 

 For real energy shifts we have found the phenomenon of 
fission of states and moving  their parts away from one another when their 
 energy levels approach one another (gradual degeneration). The exact 
 degeneracy leads to effective "annihilation" of states \cite{CZBDS}. We 
 have found that the same phenomenon occurs for two states with the 
neighbour complex eigenvalues. It is illustrated in figures 3, 4 obtained 
by using one-step SUSYQ transformation, Eqs. (\ref{trans}) - (\ref{psin}). 
 
 This enriches our quantum intuition :
 if we squeeze space localization 
of states, they become less dense on the energy scale and vice versa. The 
more broad is a potential well, the denser becomes the energy spectrum.  

It is surprising that for the infinite initial rectangular well strong 
potential perturbation slightly change  the probability 
distribution for the shifted state (small difference between absolute 
values of initial and resulting  wave functions). It is despite that the 
real and imaginary parts of the wave function are drastically changed.  
Also for all other bound states absolute values of eigenfunctions  are 
approximately conserved under transformation.  

  In the case of  IP approach  the imaginary shift of a 
ground state energy level $E$ in initial 
infinite rectangular well to $E-i$ violates  boundary conditions. 

Let us now discuss the complex transformations preserving all real 
physical eigenvalues. Again we shall consider  the  infinite 
rectangular potential model.  The complete spectrum of physical states 
corresponds to zero boundary conditions ($\psi =0$ at both walls). We can 
also consider auxiliary "non-physical" eigenvalue problem with boundary 
conditions:  $\frac{d}{dx} \psi=0$ and $\psi=0$ at different 
boundaries.  If now we add to one non-physical eigenvalue some imaginary 
part we shall get by using the  SUSYQ formalism a complex potential.  There 
is a {\it theorem of two real spectra} (see \cite{ChS} p. 401, \cite{ZS} p. 
34 and references therein) which have analog for complex-valued 
eigenvalues.  Daskalov suggested algorithms of ESM construction when {\it 
the control levers are two-spectra parameters} which determine potential 
uniquely in contrast to {\it IP approach where the complete spectral data 
are eigenvalues and spectral weights} (reduced widths).  It is possible to 
vary one eigenvalue keeping all other unchanged (in this case spectral 
weights are not conserved).  So we can get real physical spectrum identical 
to one of the initial  eigenvalue problem but with real potential 
transformed into complex one. We have checked the biorthogonality of 
complex eigenfunctions which plays a part of usual orthogonality for 
self-ajoint operators \cite{NL}. There is also a corresponding analog of 
the completeness relation.

To have a notion about possible shapes of corresponding potential
and function transformations, it is instructive to consider, as an 
especially simple example, one half of a  
potential with right-left symmetry according to its middle point at the  
origin.  {\it In this case the physical spectrum of the whole potential is 
a sum of two spectra for the half of potential corresponding to two 
different boundary conditions} at the origin ($\frac{d}{dx} \psi|_{x=0}=0$ 
or $\psi|_{x=0}=0$.  Then the known rules of transformations for 
shifting odd and even energy levels in the whole potential are 
simultaneously valid for variations of chosen levels in one of two spectra. 
 
\section{Transformation of scattering states }      

 The imaginary shift (${\cal E}>0 \to {\cal E} + i \Gamma $) of energy 
 value in the continuous spectrum of free motion by using double SUSYQ 
 transformation leads to quasi-periodic (asymptotically periodic) complex 
 potential perturbation as shown in figure 5. The infinite number of 
 the potential bumps and wells correspond to the infinite bumps  
 of the  initial state ($\sin(kx)$)  There appears quadratically integrable 
decaying state at $E + i \Gamma $ drawn by dotted line. A simple 
 explanation of this fact is an open problem. The solutions on the real 
 $E$-axis become quasi-Bloch waves.  It is interesting that the continuous 
 spectrum remains without lacunas (where solutions diverge) except for a 
 single point $E={\cal E}$.  This point can be considered as a collapsed 
 "forbidden" zone of zero width, the pinned-out point (POP).  
 Simultaneously this is the junction point of boundaries of allowed zones.  
 The solution at this point is demonstrated in figure 6.  At the neighbour 
 real energy points the solution has beats (figure 7).

 For shift ${\cal E} \to {\cal E} - i\Gamma $ in  IP-approach we have 
another result :  there appears localized (asymptotically vanishing) 
 complex potential. Waves at real energies below and above POP are 
 reflectionless but with different fluxes on both sides of the 
potential perturbation. The  parameter $\Gamma$ can serve as a control 
lever of the wave absorption.

 Energy dependence of $\Delta I(E)= I(x,E)|_{x \to 
\infty } - I(x,E)|_{x \to -\infty } $ (non-conservation of 
flux $I$) has a resonance character.  The position of the resonance peak
coincides with the real part of the quasi-bound state's eigenvalue 
 ${\bar {\cal E}}$, which resembles the case of ordinary resonances.

It would be interesting to consider a possibility to use complete sets 
of eigenstates related to non-Hermitian operators for  R-matrix 
discrete parametrization of scattering data (open problem).

It turns out that purely periodical potentials can be created by using
one-step SUSYQ transformation of free motion system (whole line), Eqs. 
(\ref{trans}) -- (\ref{psin}). As  an initial solution at 
factorisation energy ${\cal E} > 0$ we take $\psi_{0}(x,{\cal E})=
\exp(-i \sqrt{{\cal E}} x) + c \exp(i \sqrt{{\cal E}} x)$.
Potential and wave function for $c=2$ and ${\cal E}=0.5 $ are shown 
 in figures 8 and 9.
 
 \section{Conclusions}

 The 100 years old good quantum mechanics acquires additional 
attractiveness with its visualization. This gives a deeper insight into 
wave objects  (generalization of spectral design). But there still remain  
open problems, for example, the extention of the qualitative theory to 
multichannel systems and revealing a lot of physical effects, which can 
be expected due to more of degrees of freedom.

\newpage

\centerline{{\Large Figure Captions}}

\vspace*{.4cm}

Figure 1. Comparison of real   $\Delta V(x)$ (in solid) shifting down 
 the ground  state energy level ($\Delta E =-1$) in the soliton-like 
initial potential   well and imaginary part (dashed line) of potential 
perturbation $Im \Delta V(x)$ (for the same initial potential) shifting the 
initial ground state energy level ($\Delta E = i \Gamma = -i $) into 
complex energy plane.

\vspace*{.4cm}

Figure 2. The remarkable qualitative similarity of $\Delta Re 
 V(x)$ shapes    for the shifts of the first and second levels in infinite 
 rectangular well: (curve 1) $E_{1}=1 \rightarrow 1-i$; (curve 2) $ \enskip 
 E_{2}=4 \rightarrow 4-i$).  It is "paradoxical" taking into account that  
 the wave functions of the shifted states are quite different.

\vspace*{.4cm}

Figure 3. Real part (solid line) and imaginary part (dashed line) of the 
transformed soliton-like  potential when creating a new state  
with the same real part of eigenvalue $E=-1-i$ as for the neighbour 
stationary state $E=-1$.  Absolute values of the wave functions for 
both states (dotted line) are splitted in two parts and concentrated mainly 
inside the deep separated potential wells.  This resembles the tendency to 
effective "annihilation" while gradual degeneracy of the energy levels in 
the case of real energy shifts [8].

\vspace*{.4cm}

Figure 4. Real part of the transformed oscillator potential 
when creating a new state with the same real part of eigenvalue $E=1-0.1 i$
(solid line) and $E=1-0.0001 i$ (dashed line) 
as for the neighbour stationary state $E=-1$. Wave functions for both 
states are splitted as in figure 3 and localized in the  narrow potential 
wells.

\vspace*{.4cm}

 Figure 5. a) Asymptotically quasi-periodic potential 
(solid line, only real part is shown) for the energy value ${\cal E}=1$ of 
the initial continuous spectrum shifted to ${\bar {\cal E}}=1-i$.  Absolute 
value  of wave function  which became 
{\it quadratically integrable} as a result of this SUSYQ 
transformation is shown by dashed line.

\vspace*{.4cm}

 Figure 6.  Real part of the wave function at  the pinned out 
point (POP) ${\cal E} =1$ of the continuum spectrum has a characteristic 
behaviour of a solution on the boundary of spectral zones with linear 
increase of amplitude (imaginary part has an analogous shape).  In this 
specific limiting case the upper and the lower boundaries of two allowed 
zones coincide: the width of the forbidden zone  is  squeezed to zero.

\vspace*{.4cm}

Figure 7. Beats of the solution at the real energy ${\cal E}+1.1$ 
near the POP ${\cal E} =1$, see figures 5. The frequency of these beats 
decreases to zero when the energy approaches POP ${\cal E}$, see figure 
6.

\vspace*{.4cm}

Figure 8. Real (solid line) and imaginary (dashed line) parts of periodic 
 potential obtained by one-step SUSYQ transformation of free motion system 
at the point ${\cal E}=0.5$.

\vspace*{.4cm}

 Figure 9.  Real (solid line) and imaginary (dashed line) parts of periodic 
  wave function at the energy ${\cal E}=0.5 $, see figure 8.
  

\begin{thebibliography}{99 }                                  

\bibitem{ChS} K. Chadan and  P. Sabatier, {\it Inverse Problems in 
Quantum Scattering Theory }, 2nd ed. (Springer, Heidelberg, 1989).

\bibitem{CKS} F. Cooper,  A. Khare A and  U. Sukhatme, Phys. Rep. {\bf 
251}, 267 (1985).

\bibitem{PT}  J. Poshel and  E. Trubowitz, {\it Inverse Spectral Theory 
 } ( Academic Press, New York, 1987).

\bibitem{ZS}  B. N. Zakhariev and  A. A. Suzko,  {\it Direct and 
Inverse Problems } ( Springer, Heidelberg, 1990).

\bibitem{CZ1}  B. N. Zakhariev, N. A. Kostov, and E. B. Plekhanov,  
Phys. Part. \& Nucl.  {\bf 21},  384 (1990).

\bibitem{Z2}   B. N. Zakhariev,  Phys.  Part. \&  
Nucl. {\bf 23},  603 (1992). 

\bibitem{BP} V. P. Berezovoj and A. I. Pashnev, 
Sov. J. Math. Phys. {\bf 70}, 146 (1987);   Z.Phys. C {\bf  51}, 525 
(1991).
    
\bibitem{CZBDS}  V. M. Chabanov,  B. N. Zakhariev,  S. Brandt, 
H. D. Dahmen, and N. Stroh, Phys. Rev. A {\bf 52}, R3389 (1995).

\bibitem{CZS} V. M. Chabanov, B. N. Zakhariev,  and S. A. Sofianos, 
Ann. Physik {\bf 6}, 136 (1997).
 
\bibitem{CZP} V. M. Chabanov and B. N. Zakhariev, Phys. Lett.
A {\bf 255}, 123 (1999).


\bibitem{ZM} B. N. Zakhariev and M. A. Mineev, J. Mosc. Phys. Soc.
{\bf 7}, 227 (1997).

\bibitem{Suk}  C. V. Sukumar, J.  Phys.  A:  Math Gen. {\bf 20}, 
 2461 (1987), see also references therein.


\bibitem{Z} B. N. Zakhariev, {\it Lessons in Quantum Intuition}
(JINR, Dubna, 1996), (in Russian);  {\it New ABC of Quantum Mechanics} 
(UdSU, Izhevsk, 1998), (in Russian).

\bibitem{CZ3} V. M. Chabanov and B. N. Zakhariev, Phys.  Part. \&  
Nucl.  {\bf 25},  662  (1994).

\bibitem{CZ4}  V. M. Chabanov  and  B. N. Zakhariev, 
Phys. Part. \&  Nucl. {\bf 30},  111 (1999).

\bibitem{ChZ}  V. M. Chabanov  and  B. N. Zakhariev,  Inverse Problems 
 {\bf 13},  R47 (1997).

\bibitem{CZA} V. M. Chabanov, B. N. Zakhariev, and I. V. Amirkhanov,
 Ann. Phys. (NY) {\bf 285}, 1 (2000).


\bibitem{NL} M. A. Naimark,   {\it Linear Differential Operators } 
(Nauka, Moscow, 1969) (in Russian).

\bibitem{Zn} M. Znojil, J. Phys. A:  Math Gen. {\bf 33}, 4203 (2000);
{\it ibid}, 4561 (2000).

\bibitem{B} D. Baye,  J-M Sparenberg, and G. Levai, Nucl.  Phys. A 
{\bf 599}, 435 (1996).

\bibitem{A} A. A. Andrianov, F. Cannata, J-P Dedonnder, and M. V. Ioffe, 
e-print quant-ph/9806019.                                                                 


\end{thebibliography}
\end{document}